\pdfoutput=1
\documentclass[twocolumn,
               showpacs,
               preprintnumbers,
               nofootinbib,
               prd,
               superscriptaddress,
               10pt,
               notitlepage,
               aps]{revtex4-2}
               
\usepackage{graphicx,amssymb,amsmath,amsthm,amsfonts,epsfig,mathtools}
\usepackage[utf8]{inputenc}
\usepackage{graphicx}
\usepackage{dcolumn}
\usepackage{bm}
\usepackage{color}
\usepackage[dvipsnames]{xcolor}
\usepackage{hyperref}
\hypersetup{colorlinks=true, citecolor=MidnightBlue,linkcolor=CornflowerBlue, urlcolor=CornflowerBlue, linktocpage=true}
\usepackage{xfrac}
\usepackage{siunitx}
\usepackage{soul}
\newcommand{\beqn}{\begin{eqnarray}}
\newcommand{\eeqn}{\end{eqnarray}}
\newcommand{\beq}{\begin{equation}}
\newcommand{\eeq}{\end{equation}}

\def\d{\mathrm{d}}

\begin{document}

\title{Crossing the singularity of a gravitational wave collision}

\begin{abstract}
A reformulation of general relativity inspired by the Belinski–Khalatnikov–Lifshitz conjecture had been introduced by Ashtekar, Henderson and Sloan which is based on variables closely related to the basic variables of loop quantum gravity, thereby providing a way of classically analyzing singularities that may be carried over to the quantum theory. It is reasonable to expect that these variables are regular at generic spacelike singularities. This has been shown on various examples---particularly, cosmological spacetimes. In this study we extend this analysis to the singularities of gravitational wave collision spacetimes, which are the result of the mutual focusing of the two waves. We focus on two specific examples and explicitly confirm that the said variables are regular at the singularity and can be smoothly continued beyond it.
\end{abstract}

\author{Tekin Dereli}
\email{tekindereli@maltepe.edu.tr}
\affiliation{Department of Basic Sciences, Faculty of Engineering and Natural Sciences, Maltepe University, 34857 Maltepe, Istanbul, Turkey}
\affiliation{Department of Physics, Ko\c{c} University, \\
Rumelifeneri Yolu, 34450 Sariyer, Istanbul, Turkey}

\author{Ozay Gurtug}
\email{ozaygurtug@maltepe.edu.tr}
\affiliation{Department of Basic Sciences, Faculty of Engineering and Natural Sciences, Maltepe University, 34857 Maltepe, Istanbul, Turkey}

\author{K{\i}van\c{c} \.I. \"Unl\"ut\"urk}
\email{kunluturk17@ku.edu.tr}
\affiliation{Department of Physics, Ko\c{c} University, \\
Rumelifeneri Yolu, 34450 Sariyer, Istanbul, Turkey}

\date{\today}
\maketitle

\section{Introduction}

Spacetime singularities in general relativity need to be understood better for the deterministic nature of the theory. Classical tools for understanding and resolving the issue of singularities are insufficient due to the scales at which singularities are formed. Thus, quantum tools are required to cope with the analysis of singularities. For instance, test quantum wave packets obeying Klein-Gordon, Dirac and Maxwell equations may be used to probe timelike singularities in static spacetimes \cite{Horowitz1995, Helliwell2003, Gurtug2014}. In this formalism, the quantum singularity is understood as a non-unique evolution of test quantum wave packets. In dynamical spacetimes such as cosmological ones, the interior of the Schwarzschild black hole or colliding gravitational wave spacetimes, our example of interest here, a new perspective is essential for analyzing spacelike singularities. To this end, the formalism introduced by \textcite{Ashtekar2009, Ashtekar2011}, which is motivated by the Belinski–Khalatnikov–Lifshitz (BKL) conjecture \cite{Belinskii1970}, arise as a promising tool in dealing with spacelike singularities in dynamical spacetimes.

According to the BKL conjecture, close to a generic spacelike singularity, the spatial derivatives in the field equations become negligible compared to the time derivatives, turning the evolution equations approximately into ordinary differential equations, and the metric exhibits the same kind of basic chaotic behaviour, which can be accurately captured by the Bianchi models. There is now a body of numerical and analytical evidence in favor of the conjecture \cite{Weaver1998, Andersson2001, Berger2002, Garfinkle2004, Garfinkle2007, Saotome2010}, thus, it is intriguing to think about its consequences for quantum gravity. As pointed out by \textcite{Ashtekar2011}, it is possible that the BKL behaviour sets in already when the spacetime is sufficiently classical, i.e., relatively away from the Planck scale. This means that a quantization of the effective theory with ordinary differential equations could provide a qualitative understanding of quantum gravity effects close to spacelike singularities in general.

It is natural to expect that the singularities in classical general relativity are resolved in a fully quantum theory of gravity. Indeed, there is already an accumulated body of evidence in this direction, such as the ones coming from loop quantum cosmology (LQC), where the singularities of cosmological models are replaced effectively by a bounce. Apart from the FLRW metric \cite{Bojowald2001}, this has been shown also for the Bianchi I, II and IX models \cite{Ashtekar2009a, Ashtekar2009b, WilsonEwing2010}. In light of the BKL conjecture, these results suggest that generic spacelike singularities may all be resolved in LQC. However, the standard formulations of the BKL conjecture are in terms of variables not particularly suited for quantization. E.g., in the formalism introduced by \textcite{Uggla2003}, one normalizes relevant physical quantities by an appropriate power of the inverse of the trace of the extrinsic curvature $K$, which is expected to diverge at the singularity, thus giving finite quantities.

Loop quantum gravity is based on a density weighted triad formulation of general relativity, i.e., the basic configuration variables are the orthonormal triads on spacelike surfaces, scaled by the the square root of the determinant of the spatial metric $\sqrt{q}$. The determinant $q$ is expected to vanish at spacelike singularities, just like $K^{-1}$. Thus, another strategy to formulate the BKL conjecture would be to multiply divergent quantities by appropriate powers of $q$\footnote{We note that such a strategy was also considered by \textcite{Einstein1935}.}.

Based on these observations Ashtekar, Henderson and Sloan (AHS) introduced a formulation of the BKL conjecture \cite{Ashtekar2009, Ashtekar2011} based on variables closely related to the basic variables of loop quantum gravity, and therefore better suited for quantization. These variables are tensor densities involving positive powers of $q$, and therefore expected to be finite at the singularity. So far, the AHS variables have been used to study the singularities of the FLRW spacetime with a massless scalar field, Bianchi I and IX models and the part of the Schwarzschild spacetime inside the horizon \cite{valdesmeller2021}. In all of these cases, the AHS variables are regular at the singularity, and can be continued beyond it.

In this study, we extend this analysis to the singularities of colliding gravitational wave spacetimes. Colliding gravitational waves have been known to produce singularities as a result of the mutual focusing of the two waves. Though the development of singularities seem to be a generic feature of colliding gravitational waves, there also exist exceptional solutions where, instead of a singularity, a Cauchy horizon may develop in the region of interaction of the two waves \cite{Bell1974, Chandrasekhar1986, Ferrari1988}. However, it has been shown that these horizons are unstable in the full nonlinear theory against small but generic plane-symmetric perturbations of the incoming waves, and thus, spacelike curvature singularities are developed in the interaction region \cite{Yurtsever1987}. Furthermore, it has been shown that close to the singularity, these spacetimes asymptotically approach the Kasner metric \cite{Yurtsever1988}, i.e. a Bianchi type-I metric, in accordance with the BKL conjecture.

Colliding gravitational wave spacetimes therefore provide interesting models for investigating singularities, which are relatively unexplored. Here, we shall calculate the AHS variables for two basic examples of colliding wave solutions and confirm explicitly that the AHS variables are regular at the singularity and can be continued smoothly through the singularity.

The paper is organized as follows. In Section~\ref{sec: colliding GW review}, we briefly review the basics of colliding gravitational waves. In Section~\ref{sec: AHS variables} we summarize the triad formulation of general relativity and reintroduce the AHS variables. Then, we demonstrate the application of the variables using the example of the Schwarzschild spacetime. In Section~\ref{sec: application to gw} we apply the formalism to the two above--mentioned colliding wave examples. Finally, in Section~\ref{sec: discussion} we discuss the implications of our results and the further questions that need to be addressed. Throughout the paper we use the mostly plus metric signature $(-,+,+,+)$ and $c=1$.

\section{An overview of colliding gravitational waves}
\label{sec: colliding GW review}

\begin{figure}
    \centering
    \includegraphics[width=0.85\columnwidth]{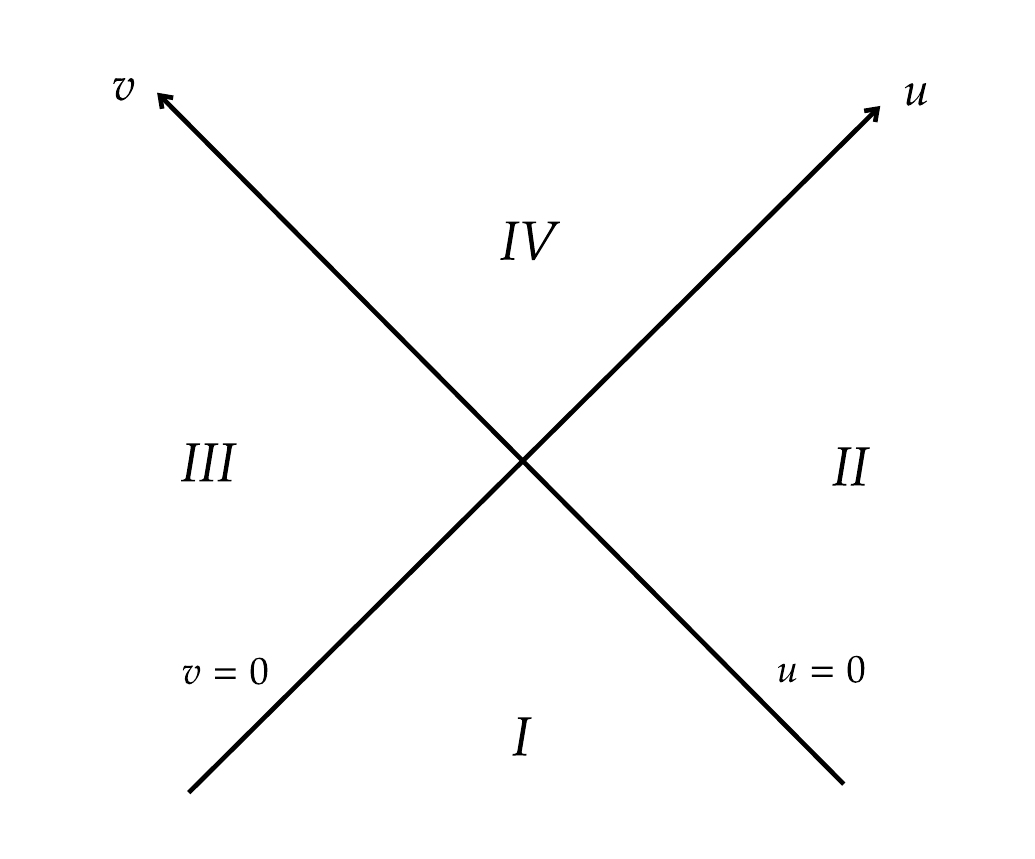}
    \caption{General structure of a colliding wave spacetime. Regions II and III contain the incoming waves, possibly including impulsive waves on the boundaries $u=0$ and $v=0$. There are no waves in region I, which is therefore described by the background metric. The waves start interacting after the point $u=v=0$, so that the metric in region IV is determined by the interaction of the waves. Generically, the interaction results in a curvature singularity in region IV.}
    \label{fig: 4 regions}
\end{figure}

Before introducing the colliding gravitational wave solutions we are going to work with, it would be appropriate here to briefly review the fundamental concepts of the subject. When studying such collision solutions, it is convenient to divide spacetime into four regions, as shown in Fig.~\ref{fig: 4 regions}, with two dimensions suppressed. In a typical collision scenario, one has two approaching waves coming from infinity, so that regions II and III contain the incoming waves, whereas region I contains no waves and is therefore described by the---typically flat---background metric. After the point $u=v=0$, the two waves interact, and region IV is described by the interaction metric, which is to be determined.

In Rosen coordinates, the interaction of linearly polarized gravitational waves, possibly coupled with matter waves, is described \cite{Griffiths1991} by a metric of the form
\begin{equation}
g = -2e^{-M}\d u \d v + e^{-U} \left(e^V  \d x^2 + e^{-V}\d y^2\right),
\label{eq: metric general null form}
\end{equation}
where the metric functions $M=M(u,v)$, $U=U(u,v)$ and $V=V(u,v)$ are functions of the null coordinates $u$ and $v$. These functions will assume different forms in each region, which must satisfy certain junction conditions at each boundary separating two regions.

A convenient way to examine the singularity of a colliding gravitational wave spacetime is provided by the Newman-Penrose (NP) formalism through the Weyl-NP scalars $\Psi_i$. For instance, if $\Psi_1=\Psi_3=0$, as will be the case in the examples we look at, the two polynomial curvature invariants $I$ and $J$ \cite{Penrose2003} can be written as
\begin{equation}
I = 2\Psi_0\Psi_4 + 6\Psi_2^2, \quad J=6\left(\Psi_0\Psi_4 - \Psi_2^2\right)\Psi_2.
\end{equation}
Note in particular that if $\Psi_2$ becomes unbounded, either $I$ or $J$ must also be unbounded, hence, showing that $\Psi_2$ diverges is a sufficient proof of a curvature singularity in this case.

The metric in Eq.~\eqref{eq: metric general null form} suggests a natural choice for the complex null tetrad:
\begin{subequations}
\begin{align}
l_{\alpha} & =e^{-M/2}\delta^{u}_{\alpha},\\
n_{\alpha} & =e^{-M/2}\delta^{v}_{\alpha},\\
m_{\alpha} & =\frac{e^{-U/2}}{\sqrt{2}}\left(e^{V/2}\delta^{x}_{\alpha} + ie^{-V/2}\delta^{y}_{\alpha}\right),\\
\bar{m}_{\alpha} & =\frac{e^{-U/2}}{\sqrt{2}}\left(e^{V/2}\delta^{x}_{\alpha} - ie^{-V/2}\delta^{y}_{\alpha}\right).
\end{align}
\end{subequations}
Working with this, it is straightforward to see that for the metric~\eqref{eq: metric general null form}, we have $\Psi_1=\Psi_3=0$ and
\begin{subequations}
\begin{align}
\Psi_0 &= -\frac{1}{2}e^M \left[\left(M_v - U_v\right)V_v + V_{vv}\right], \\
\Psi_2 &= \frac{1}{6}e^M \left(M_{uv} - U_{uv} + V_u V_v\right), \\
\Psi_4 &= -\frac{1}{2}e^M \left[\left(M_u - U_u\right)V_u + V_{uu}\right],
\end{align}
\end{subequations}
where the subscript denotes a partial derivative with respect to the coordinate.

Typically, the Weyl-NP scalars $\Psi_4=\Psi_4(u)$ in region II and $\Psi_0=\Psi_0(v)$ in region III represent the incoming gravitational waves that participate in the collision. In the interaction region these become functions of both of the null coordinates; $\Psi_0=\Psi_0(u,v)$ and $\Psi_4=\Psi_4(u,v)$. In addition, a finite Coulomb component $\Psi_2=\Psi_2(u,v)$ also develops as a result of the nonlinear interaction.

On the other hand, the Ricci-NP scalars $\Phi_{ij}$ represent the electromagnetic or other matter fields participating in the collision, e.g., $\Phi_{00}(v)$ and $\Phi_{22}(u)$ represent electromagnetic wave components in the incoming regions. In the interaction region, $\Phi_{00}(u,v)$ and $\Phi_{22}(u,v)$ are necessarily nonzero and $\Phi_{02}(u,v)$ develops as a result of the nonlinear interaction.

In this study, our focus will be on two specific colliding gravitational wave solutions. The first is the Khan-Penrose solution, which was the first solution discovered in this context and which describes the collision of two impulsive plane symmetric gravitational waves. The second is the collision of a plane electromagnetic wave with plane impulsive gravitational waves accompanied by shock gravitational waves. The common point in each solution is the existence of a spacelike curvature singularity in the interaction region.

Since our focus in this study is the singularity structures of these spacetimes, which are located in the interaction region, we shall not be interested in the solutions in regions I, II and III, and only consider region IV.

\subsection{Khan-Penrose solution}
The first exact solution that considers the collision of two gravitational waves was discovered by \textcite{Khan1971}, where the colliding waves are linearly polarized, impulsive gravitational waves described by $\Psi_0=\delta(v)$ and $\Psi_4=\delta(u)$.

\begin{figure}
    \centering
    \includegraphics[width=\columnwidth]{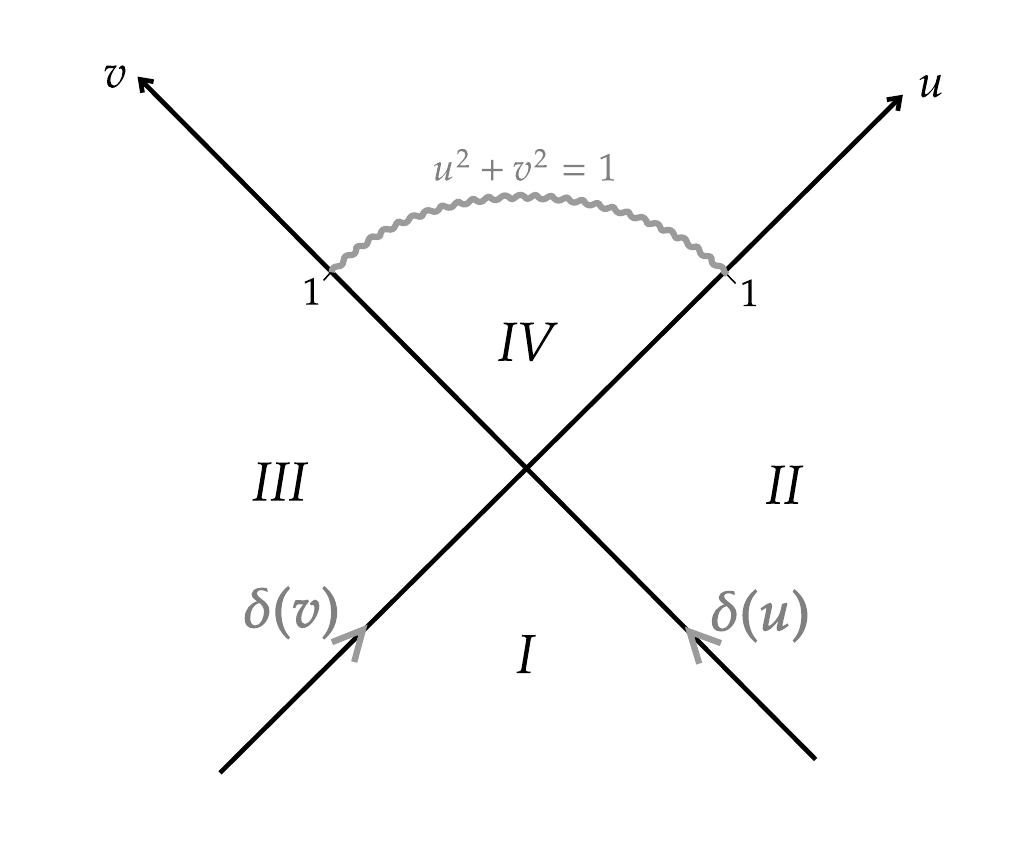}
    \caption{Structure of the Khan-Penrose spacetime. The incoming Dirac-$\delta$ waves start interacting after the point $u=v=0$, and eventually create a singularity on the surface $u^2 + v^2 = 1$. The part of region IV below the singularity is described by the metric~\eqref{eq: metric general null form} with the metric functions given by Eq.~\eqref{eq: khan-penrose metric functions}.}
    \label{fig: khan-penrose diagram}
\end{figure}

The mutual focusing of the two waves eventually produces a spacelike singularity. The interaction region of the Khan-Penrose spacetime is described by the metric~\eqref{eq: metric general null form}, where the metric functions are given by
\begin{subequations}
\begin{align}
e^{-U} &= 1-u^2-v^2, \\
e^{-V} &= \frac{1+u\sqrt{1-v^2}+v\sqrt{1-u^2}}{1-u\sqrt{1-v^2}-v\sqrt{1-u^2}}, \\
e^{-M} &= \frac{\left(1-u^2-v^2\right)^{3/2}}{w^2\sqrt{1-u^2}\sqrt{1-v^2}},
\end{align}
\label{eq: khan-penrose metric functions}
\end{subequations}
with $w=uv+\sqrt{1-u^2}\sqrt{1-v^2}$. In this region, we get
\begin{equation}
\Psi_2 = e^{7U/2} w^2 \left(w^2 - uv\sqrt{1-u^2}\sqrt{1-v^2}\right).
\label{eq: psi2 of khan-penrose}
\end{equation}
Note that on the spacelike surface $e^{-U}=0$, i.e.,
\begin{equation}
    u^2 + v^2 = 1,
\end{equation}
the term inside the parentheses in Eq.~\eqref{eq: psi2 of khan-penrose} is finite. Thus, $\Psi_2$ diverges as ${\sim}\,e^{7U/2}$, showing that the solution becomes singular. The singular surface of the Khan-Penrose spacetime is illustrated in Fig.~\ref{fig: khan-penrose diagram}.

\subsection{Colliding electromagnetic and gravitational waves}
The second example we shall look at is the collision of electromagnetic waves with plane symmetric impulsive gravitational waves accompanied by shock gravitational waves. This solution was obtained by \textcite{Gurtug2006} together with an additional scalar field. Here, we shall take the vanishing scalar field limit of this solution for simplicity. The incoming gravitational wave is described by $\Psi_4 = \delta(u) - 3(1+\sin u)^{-3}$ in region II and the incoming electromagnetic wave may be described by the Ricci-NP scalar $\Phi_{00}=1$ in region III.

The interaction region is described again by Eq.~\eqref{eq: metric general null form}, this time with the metric functions
\begin{subequations}
\label{eq: metric component functions em-gw}
\begin{align}
e^{-U} & =\cos^{2}u+\cos^{2}v-1, \\
e^{-V} & =\left(1+\sin u\right)^{2}, \\
e^{-M} & =\cos u\cos v\,e^{-V+U/2},
\end{align}
\end{subequations}
and the electromagnetic potential $A_\mu = A \delta^y_\mu$, where
\begin{equation}
A = \sqrt{2} \sin v \left(1 + \sin u \right).
\end{equation}

\begin{figure}
    \centering
    \includegraphics[width=\columnwidth]{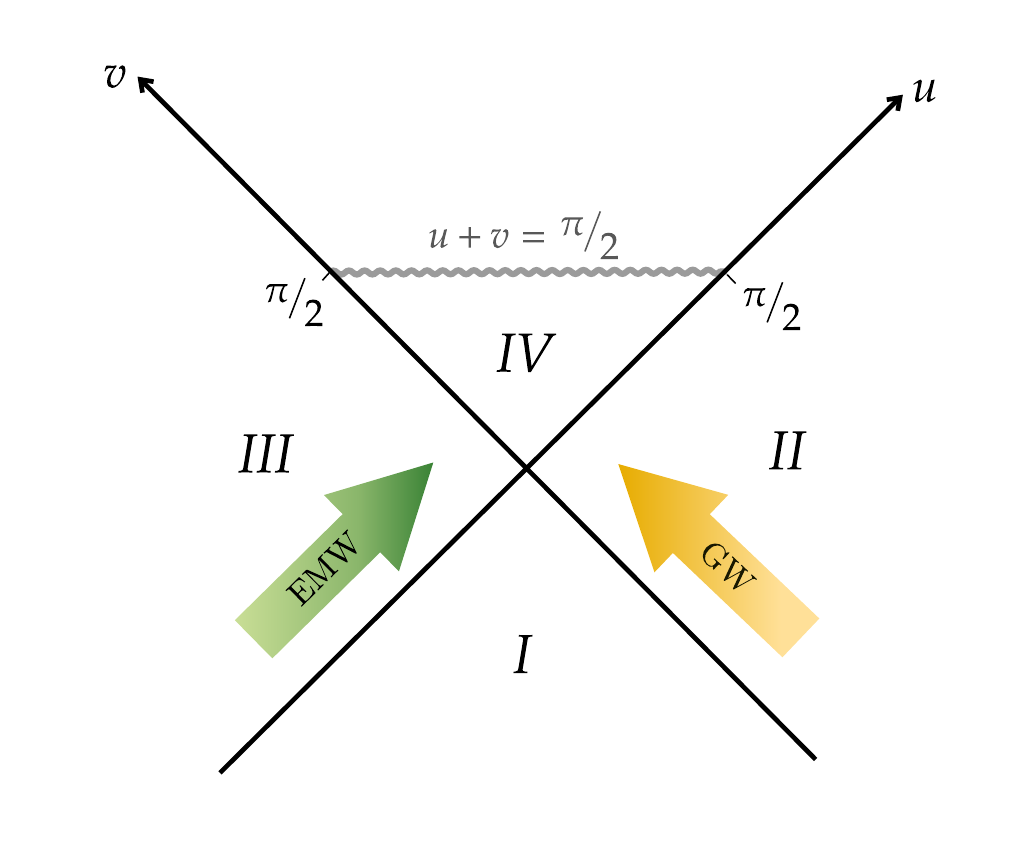}
    \caption{Structure of the gravitational-electromagnetic wave collision spacetime. The gravitational waves (GW) coming from region II and the electromagnetic waves (EMW) coming from region III start interacting after the point $u=v=0$, and the interaction results in a singularity on the surface $u + v = \pi/2$. The part of region IV below the singularity is described by the metric~\eqref{eq: metric general null form} with the metric functions given by Eq.~\eqref{eq: metric component functions em-gw}.}
    \label{fig: gw-em diagram}
\end{figure}

For this metric we have
\begin{equation}
\Psi_2 = -e^{3U/2} \frac{\sin u \sin v}{\left(1+\sin u\right)^2}.
\end{equation}
Looking at $\Psi_2$, we see again that on the surface $e^{-U}=0$, i.e.,
\begin{equation}
    \cos^2 u + \cos^2 v = 1,
\label{eq: gw-em singularity condition}
\end{equation}
we have a spacelike singularity. Note that Eq.~\eqref{eq: gw-em singularity condition} is satisfied for $u+v = \pi/2$. This is illustrated in Fig.~\ref{fig: gw-em diagram}.

\section{Ashtekar-Henderson-Sloan variables}
\label{sec: AHS variables}

Motivated by the BKL conjecture, a set of variables was introduced by \textcite{Ashtekar2009, Ashtekar2011}, which provides a way of analyzing spacelike singularities in a purely classical setting. These variables are based on the formulation of general relativity in terms of a density weighted triad, which is also the starting point of loop quantum gravity. In this section, we shall review the triad formulation of general relativity and the Ashtekar-Henderson-Sloan (AHS) variables. We shall then summarize the application of these variables to the Schwarzschild singularity as an illustration of the idea, which will also help us introduce some of the ideas we are going to use in the next section.

We consider spacetimes of the form $\mathcal{M}=\mathbb{R}\times M$, where $M$ is a three-dimensional manifold, and foliate $\mathcal{M}$ into spacelike slices $\Sigma_{t}$. Let $q_{ab}$ denote the spatial metric induced on these slices. A density weighted triad $\tilde{E}_{i}^{a}$ on $\Sigma_{t}$ is defined by
\begin{equation}
\tilde{E}_{i}^{a}\tilde{E}^{bi}=\tilde{\tilde{q}}q^{ab},
\label{eq: triad and metric}
\end{equation}
where $\tilde{\tilde{q}}$ denotes the determinant of $q_{ab}$ and the tilde marks the density weight; a tilde above a variable means a density weight of $+1$. The letters $a,b,c,\dots$ denote the spatial indices and $i,j,k,\dots$ denote internal SO(3) indices. The internal indices can be raised/lowered freely using $\delta_{ij}$, while the spatial indices are raised/lowered using the spatial metric $q_{ab}$.

The triad determines a unique connection $\Gamma^i_a$ such that
\begin{equation}
D_a \tilde{E}^b_i - \epsilon^j_{\phantom{j}ik}\Gamma^k_a\tilde{E}^b_j = 0,
\end{equation}
where $D_a$ is the Levi-Civita connection of the spatial metric $q_{ab}$. We should emphasize that $D_a$ acts only on spacetime indices whereas $\Gamma^i_a$ acts only on the internal indices.

The variable canonical to $\tilde{E}_{i}^{a}$ is $K_{a}^{i}$, which, on solutions, is related to the extrinsic curvature $K_{ab}$ of $\Sigma_{t}$ by
\begin{equation}
K_{ab}=K_{(a}^{i}e_{b)i}.\label{eq: conj momentum and extrinsic curvature}
\end{equation}
The standard ADM variables $(q_{ab},\tilde{P}^{ab})$, where $\tilde{P}^{ab} = (16\pi G)^{-1}\widetilde{\sqrt{q}} (K^{ab} -K_{\phantom{c}c}^{c}q^{ab})$ can be found using Eqs. (\ref{eq: triad and metric}) and (\ref{eq: conj momentum and extrinsic curvature}).

The AHS variables \cite{Ashtekar2009, Ashtekar2011} are defined as\footnote{Note that, due to the placement of the indices, our convention for $\Gamma_{a}^{i}$ and consequently for $\tilde{C}_{i}^{\phantom{i}j}$ differs from that of \textcite{Ashtekar2009, Ashtekar2011} by a sign.}
\begin{subequations}
\label{eq: ahs definition}
\begin{align}
\tilde{P}_{i}^{\phantom{i}j} & = \tilde{E}_{i}^{a}K_{a}^{j}-\delta_{i}^{j}\tilde{E}_{k}^{a}K_{a}^{k}, \\
\tilde{C}_{i}^{\phantom{i}j} & =\tilde{E}_{i}^{a}\Gamma_{a}^{j}-\delta_{i}^{j}\tilde{E}_{k}^{a}\Gamma_{a}^{k}.
\end{align}
\end{subequations}
Furthermore, a derivative operator $\tilde{D}_{i}$ is defined as $\tilde{D}_{i}=\tilde{E}_{i}^{a}D_{a}$. Note that these carry internal indices only; they are spacetime scalars (with density weight $+1$). Physically, the $\tilde{P}^{ij}$ are related directly to the canonical ADM momenta $\tilde{P}^{ab}$, whereas the $\tilde{C}^{ij}$ encode information about the $\tilde{D}_i$ derivatives of the triad $\tilde{E}_{i}^{a}$ \cite{Ashtekar2011}. As mentioned previously, the determinant $q$ is expected to vanish at the singularity and therefore the AHS variables, being densities, are expected to be finite there. In what follows, we shall omit the tildes for simplicity. Thus, each of $E_{i}^{a}$, $C_{i}^{\phantom{i}j}$, $P_{i}^{\phantom{i}j}$ and $D_{i}$ carries a density weight of $+1$.

One can write down the equations of motion for the variables $P_{i}^{\phantom{i}j}$, $C_{i}^{\phantom{i}j}$ and the operator $D_{i}$ by taking the Poisson brackets with the Hamiltonian. We shall not use the equations of motion directly, and therefore do not repeat them here (see \textcite{Ashtekar2011} for the full equations of motion). The important thing to note, however, is that both the constraints (hence the Hamiltonian) and the time derivative of the triplet $(P_{i}^{\phantom{i}j}, C_{i}^{\phantom{i}j}, D_{i})$ can be written entirely in terms of the triplet itself (and the Lagrange multipliers $N,N^{i}$ and $\Lambda^{i}$), i.e., without reference to $(E_{i}^{a},K_{a}^{i})$.

These variables can therefore be used as follows. On an initial time slice, one can construct the triplet $(P_{i}^{\phantom{i}j},C_{i}^{\phantom{i}j},D_{i})$ from the pair $(E_{i}^{a},K_{a}^{i})$ and then evolve the triplet $(P_{i}^{\phantom{i}j},C_{i}^{\phantom{i}j},D_{i})$ without referring to $(E_{i}^{a},K_{a}^{i})$. As we have mentioned, these variables are expected to be finite at the singularity, where the metric formulation fails, hence one can continue time evolution past the singularity. Once we cross the singularity, we can, in principle, reconstruct the triad $E_{i}^{a}$ by solving an ordinary differential equation.

Note that, since the derivative operator $D_i$ has a density weight of +1, we expect the $D_i$ derivatives of regular quantities to vanish at the singularity. Thus, the $D_i$ derivatives can be interpreted as the spatial derivatives that should be negligible according to the BKL conjecture. This allows one to make a precise formulation of the BKL conjecture, where the time derivative that should dominate is the Lie derivative in the time direction of the foliation, and the spatial derivatives that should be negligible are the $D_i$ derivatives, see \textcite{Ashtekar2009, Ashtekar2011}.

\subsection{Application to Schwarzschild spacetime}
In order to demonstrate the use of AHS variables with a simple, analytical example, we shall summarize their application to the Schwarzschild singularity, which has been done by \textcite{valdesmeller2021}. We start with the metric
\begin{equation}
    g = -T(\tau)\d\tau^2 + X(\tau)\d x^2 + A(\tau)\left(\d\theta^2 + \sin^2\theta\d\phi^2\right),
\label{eq: schwarzschild quadratic form}
\end{equation}
where
\begin{equation}
    T=\frac{4\tau^4}{2M-\tau^2}, \quad X=\frac{2M-\tau^2}{\tau^2},\quad A=\tau^4.
\end{equation}
The portion of this spacetime given by $\tau\in(-\sqrt{2M},0)$ is precisely the interior of a Schwarzschild black hole, as can be seen by the coordinate transformation
\begin{equation}
    t=x, \quad r=\tau^2.
\end{equation}

\begin{figure}
    \centering
    \includegraphics[width=0.75\columnwidth]{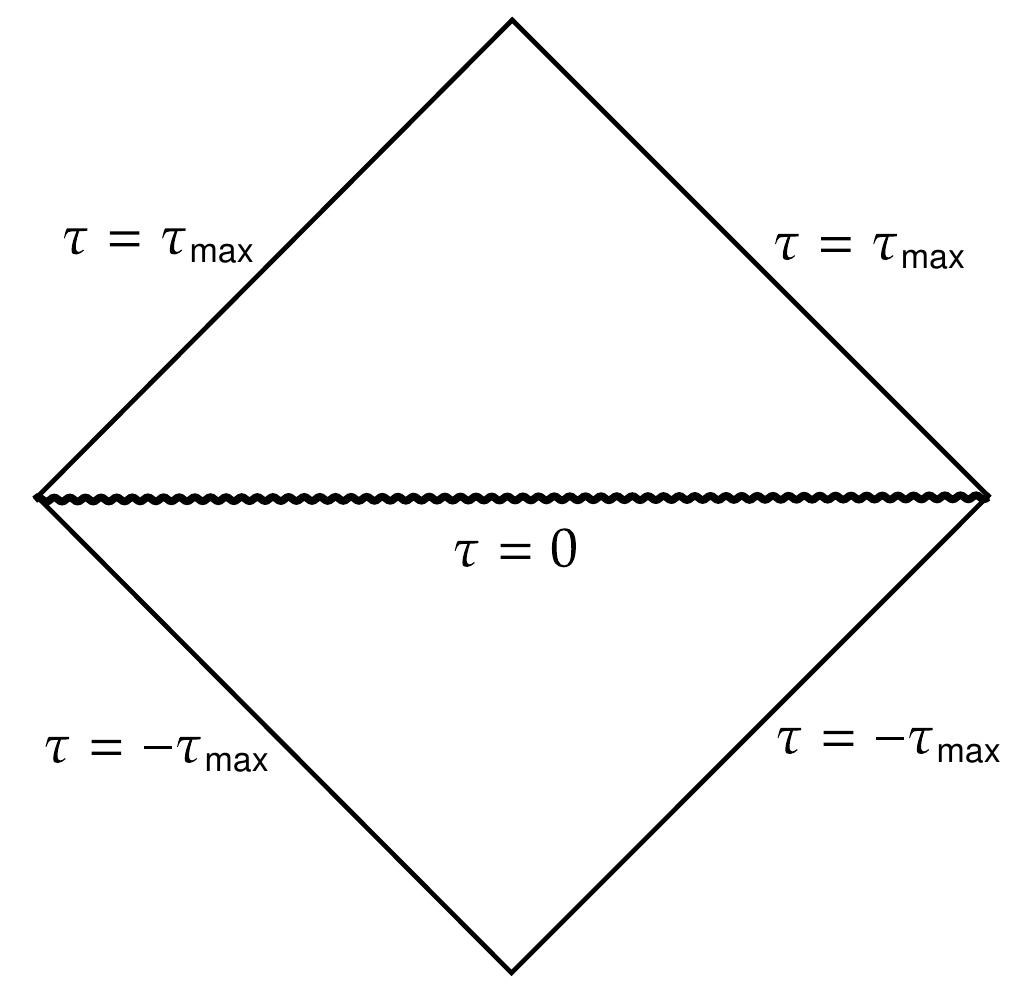}
    \caption{Spacetime diagram of the metric~\eqref{eq: schwarzschild quadratic form}. We have a well-defined metric tensor in the two regions $\tau<0$ and $\tau>0$, which are separated by the singularity, $\tau=0$. Here, $\tau_\text{max}=\sqrt{2M}$.}
    \label{fig: double copy diagram}
\end{figure}

However, the spacetime in Eq.~\eqref{eq: schwarzschild quadratic form} also includes the part $\tau\in(0,\sqrt{2M})$, which is a time--reversed copy of the black hole interior, i.e., a white hole interior (see Fig.~\ref{fig: double copy diagram}), since the transformation from $\tau$ to $r$ is a two-to-one mapping. As discussed by \textcite{DAmbrosio2018}, such a spacetime with two copies of the Schwarzschild metric may actually provide an effective description of an underlying quantum geometry, where there is a high--curvature cutoff that causes the geometry to ``bounce back" instead of becoming singular, as in the Big Bounce scenario from LQC.

The metric~\eqref{eq: schwarzschild quadratic form} has singularities at $\tau=\pm\sqrt{2M}$, which, of course, correspond to the horizon of the Schwarzschild black hole, and are merely coordinate singularities.

It also has a singularity at $\tau=0$, which is the central singularity of the Schwarzschild black hole. At both sides of the singularity, i.e. at $\tau\in(-\sqrt{2M},0)\cup(0,\sqrt{2M})$, we have a well defined metric tensor. However, in the standard metric formulation, one cannot evolve the metric from one side to the other, past the singularity. This brings up the question, can there be a formulation of the gravitational field equations which does not face the same problem at the singularity?

It is always possible in principle that a differential equation is stated in terms of ``wrong" variables, and the same equation can be stated in terms of other variables which has well-behaved solutions. To quote the simple example from \textcite{DAmbrosio2018}, the differential equation $y\ddot{y} - 2\dot{y}^2 - y^2 = 0$ has the solution $y=1/\sin t$, which is singular at $t=0$. Formulated in terms of $x=y^{-1}$, however, the equation is simply the harmonic oscillator equation $\ddot{x} + x = 0$ and the corresponding solution $x=\sin t$ is regular. The idea of the AHS variables is that, although the metric formulation fails at the singularity, the AHS variables can still remain finite and provide a well-behaved description of the gravitational field.

Now, we can calculate the AHS variables Eq.~\eqref{eq: ahs definition} for the metric in Eq.~\eqref{eq: schwarzschild quadratic form}. The spatial metric on the $\tau=\text{const}$ slices reads as
\begin{equation}
    q = X(\tau)\d x^2 + A(\tau)\left(\d\theta^2 + \sin^2\theta\d\phi^2\right).
\end{equation}
We work with the density weighted triad
\begin{equation}
E_1^a = A\sin\theta\delta_x^a, \quad E_2^a=\sqrt{XA}\sin\theta\delta_\theta^a, \quad E_3^a=\sqrt{XA}\delta_\phi^a.
\end{equation}
Then, the only non-vanishing AHS variables are \cite{valdesmeller2021}
\begin{subequations}
\begin{align}
P^{11} &= 2P^{22} = 2P^{33} = -2\sin\theta\left(2M-\tau^2\right), \\
C^{31} &= -\cos\theta\ \tau\sqrt{2M-\tau^2}.
\end{align}
\end{subequations}
Hence we see that the AHS variables are indeed finite at $\tau=0$, and can be continued from one side of the singularity to the other.

\section{Singularities of gravitational wave collisions}
\label{sec: application to gw}

In this section, we shall analyze the behaviour of the AHS variables at the singularities of the Khan-Penrose metric and the gravitational-electromagnetic wave collision solution introduced in Sec.~\ref{sec: colliding GW review}. This requires a time-space split of the spacetime, thus, it is suitable to go from the null coordinates in Eq.~\eqref{eq: metric general null form} to a pair of timelike and spacelike coordinates. We therefore define $t = (u+v)/\sqrt{2}$ and $z = (u-v)/\sqrt{2}$. With these, Eq.~\eqref{eq: metric general null form} becomes
\begin{equation}
g=e^{-M}\left(-\d t^{2}+\d z^{2}\right)+e^{-U}\left(e^{V}\d x^{2}+e^{-V}\d y^{2}\right),
\label{eq: metric general tl-sl form}
\end{equation}
which will be our starting point in what follows. Note that in these coordinates, the singularity of the Khan-Penrose metric is located at $e^{-U} = 1-t^2-z^2 = 0$, whereas the singularity of the the second metric is at a constant--$t$ surface, $t=2^{-3/2}\pi$.

\subsection{Khan-Penrose metric}

To calculate the AHS variables for the Khan-Penrose metric, we want to choose such a foliation that the singular surface $e^{-U}=0$ is a limiting case of the spacelike slices. In other words, we want to introduce a timelike variable $\tau$, such that the singularity corresponds to a level surface, conveniently chosen to be $\tau=0$. In addition, just as was done in the Schwarzschild case, we let $\tau$ enter quadratically so that we get a spacetime with two copies of the interaction region connected by the singularity. We therefore define
\begin{equation}
    \tau^2 = 1 - t^2 - z^2,
\end{equation}
whose upper and lower limits for a fixed $z$ are $\pm\sqrt{1-2z^2}$ in the region we are considering. With this, the metric~\eqref{eq: metric general tl-sl form} becomes
\begin{align}
g &= \frac{e^{-M}}{\xi+z^{2}}\left(-\tau^{2}\d\tau^{2} - 2\tau z\d\tau\d z +\xi\d z^{2}\right)\nonumber\\
&\quad\quad\quad\quad\quad\quad\quad\quad\quad  + e^{-U}\left(e^{V}\d x^{2}+e^{-V}\d y^{2}\right),
\label{eq: khan-penrose quadratic t}
\end{align}
where we introduced the shorthand notation $\xi=1-\tau^{2}-2z^{2}$. Just as in the Schwarzschild case, here we have two copies of the interaction region, one of them being the time reversed copy of the other, and the two are separated by the singularity $\tau=0$.

We now turn to the calculation of the AHS variables for the metric
(\ref{eq: khan-penrose quadratic t}). We shall work on the spatial surfaces of constant
$\tau$, so that the spatial metric reads
\begin{equation}
\label{eq: khan-penrose spatial metric}
q = e^{-M}\frac{\xi}{\xi+z^{2}}\d z^{2}+e^{-U}\left(e^{V}\d x^{2}+e^{-V}\d y^{2}\right),
\end{equation}
and we work with the density--weighted triad
\begin{subequations}
\begin{align}
E_{1}^a & =e^{-U}\delta_{z}^a,\\
E_{2}^a & =e^{-\left(M+U+V\right)/2}\sqrt{\frac{\xi}{\xi+z^{2}}}\delta_{x}^a,\\
E_{3}^a & =e^{-\left(M+U-V\right)/2}\sqrt{\frac{\xi}{\xi+z^{2}}}\delta_{y}^a.
\end{align}
\end{subequations}
Then, the only non--vanishing AHS variables (Eq. (\ref{eq: ahs definition})) are
\begin{subequations}
\begin{align}
P^{11} & =-\frac{e^{-U}}{\tau\sqrt{\xi+z^{2}}}\mathcal{D}U,\\
P^{22} & =-\frac{e^{-U}}{2\tau\sqrt{\xi+z^{2}}}\left[\mathcal{D}\left(M+U+V\right)-2\tau\frac{1-\tau^{2}}{\xi}\right], \label{eq: kp P22}\\
P^{33} & =-\frac{e^{-U}}{2\tau\sqrt{\xi+z^{2}}}\left[\mathcal{D}\left(M+U-V\right)-2\tau\frac{1-\tau^{2}}{\xi}\right], \label{eq: kp P33}\\
C^{23} & =\frac{1}{2}e^{-U}\left(U_{z}-V_{z}\right),\\
C^{32} & =-\frac{1}{2}e^{-U}\left(U_{z}+V_{z}\right),
\end{align}
\label{eq: AHS for khan-penrose}
\end{subequations}
where we have defined the shorthand operator notation $\mathcal{D}=\xi\partial_{\tau}+\tau z\partial_{z}$.

We now want to show that all of these variables are regular at the singularity. To this end, note that for any function $f$ we have
\begin{equation}
\mathcal{D}f=-\frac{\tau}{\sqrt{2}}\sqrt{\xi+z^{2}}\left(f_{u}+f_{v}\right)+\frac{\tau z}{\sqrt{2}}\left(f_{u}-f_{v}\right),
\end{equation}
and
\begin{equation}
f_{z}=\frac{1}{\sqrt{2}}\left(f_{u}-f_{v}\right)-\frac{z}{\sqrt{2}}\frac{1}{\sqrt{\xi+z^{2}}}\left(f_{u}+f_{v}\right).
\end{equation}
Therefore, the question whether the AHS variables $P^{ij}$ and $C^{ij}$ are regular at $\tau=0$ is reduced to whether $e^{-U}f_{u,v}$ are regular, where $f$ is any of the functions $U,V$ and $M$.

This is easily seen to be the case for $f=U$, for which we have $\mathcal{D}U=-2\xi\tau^{-1}$ and $U_z=0$.

For $M$, we get
\begin{equation}
M_u = 3u e^U - \frac{u}{1-u^2} - 2\frac{u\sqrt{1-v^2} - v\sqrt{1-u^2}}{w\sqrt{1-u^2}}.
\label{eq: khan-penrose M_u}
\end{equation}
Note that the second and third terms on the right hand side of Eq.~\eqref{eq: khan-penrose M_u} are regular everywhere in the region $0<u,v<1$. Thus, as $\tau\to 0$ we get
\begin{equation}
e^{-U}M_u \approx 3u = \frac{3}{\sqrt{2}} \left(\sqrt{1-\tau^2-z^2} + z\right),
\end{equation}
which is also regular. Since $M$ and $U$ are symmetric functions of $u$ and $v$, the same holds for $e^{-U}M_v$.

\begin{figure}
    \centering
    \includegraphics[width=\columnwidth]{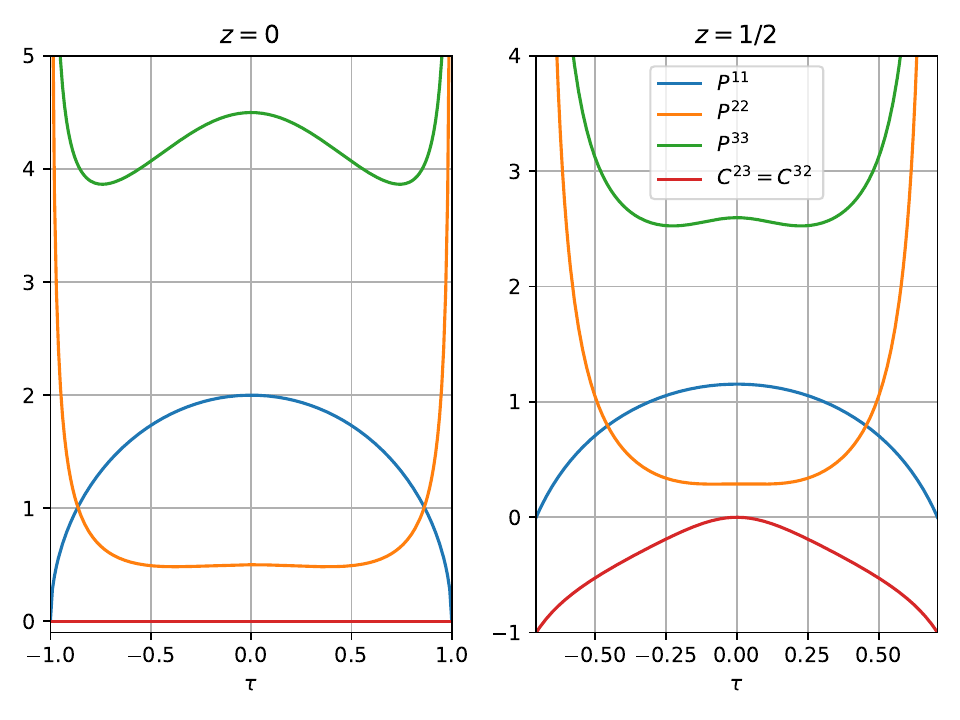}
    \caption{Non-vanishing AHS variables for the Khan-Penrose metric at two different points of space, $z=0$ (left) and $z=1/2$ (right). For this metric we always have $C^{23}=C^{32}$, which exactly vanish at $z=0$. It can be seen that all of the variables are regular at the singularity. We note that $P^{22}$ and $P^{33}$ diverge at the boundaries of our region, $\tau=\pm\sqrt{1-2z^2}$. As explained in the text, however, this does not signal a physical breakdown since the foliation we have used becomes degenerate at the boundaries.}
    \label{fig: khan-penrose ahs behav}
\end{figure}

Lastly, for $V$ we have
\begin{equation}
e^{-U}V_u = -\frac{2}{\sqrt{1-u^2}} \frac{\left(u+\sqrt{1-v^2}\right) \left(v+\sqrt{1-u^2}\right)}{1 + u\sqrt{1-v^2} + v\sqrt{1-u^2}},
\end{equation}
which is regular as well. Again, by symmetry, the same is true for $e^{-U}V_v$. Thus, all of the AHS variables in Eq.~\eqref{eq: AHS for khan-penrose} are regular throughout the region in question, including the singularity $\tau=0$, and they can be evolved smoothly from one side of the singularity to the other. Fig.~\ref{fig: khan-penrose ahs behav} shows the AHS variables for the Khan-Penrose metric as functions of $\tau$ at two sample points.

We note that $P^{22}$ and $P^{33}$ diverge at the boundaries of our region, $\tau=\pm\sqrt{1-2z^2}$, due to the $\xi^{-1}$ terms inside the square brackets in Eqs.~\eqref{eq: kp P22} and \eqref{eq: kp P33}. This does not imply a physical breakdown, however, because the boundaries are the null surfaces $u=0$ and $v=0$, or $t=|z|$, i.e., in the limit $\xi\to 0$ the foliation we have used becomes not spacelike, but null, which is also apparent from Eq.~\eqref{eq: khan-penrose spatial metric}. If one wants to use the foliation we have used to calculate and evolve the AHS variables, one simply needs to start somewhere in the interior, where the $\tau=\text{const}$ surfaces are spacelike and evolve from there.

It is easy to see that both the $C^{ij}$ and the triad $E^a_i$ vanish as one approaches the singularity, while the $P^{ij}$ assume finite values. This is just like the case for the Kasner metric \cite{Ashtekar2011}, which is what we would expect, since, as mentioned previously, the type of gravitational wave collision spacetimes we investigate here approach the Kasner metric close to the singularity, in accordance with the BKL conjecture.

\subsection{Gravitational-electromagnetic wave collision}

We now turn to the gravitational-electromagnetic wave collision solution. Similarly to the previous cases, we define a timelike variable $\tau$ as
\begin{equation}
    \tau^2 = \frac{\pi}{2\sqrt{2}} - t,
\end{equation}
so that the singularity occurs at $\tau=0$. With this definition, Eq.~\eqref{eq: metric general tl-sl form} becomes
\begin{equation}
g =e^{-M}\left(-4\tau^{2}\d\tau^{2}+\d z^{2}\right)+e^{-U}\left(e^{V}\d x^{2}+e^{-V}\d y^{2}\right).\label{eq: metric}
\end{equation}

\begin{figure}
    \centering
    \includegraphics[width=\columnwidth]{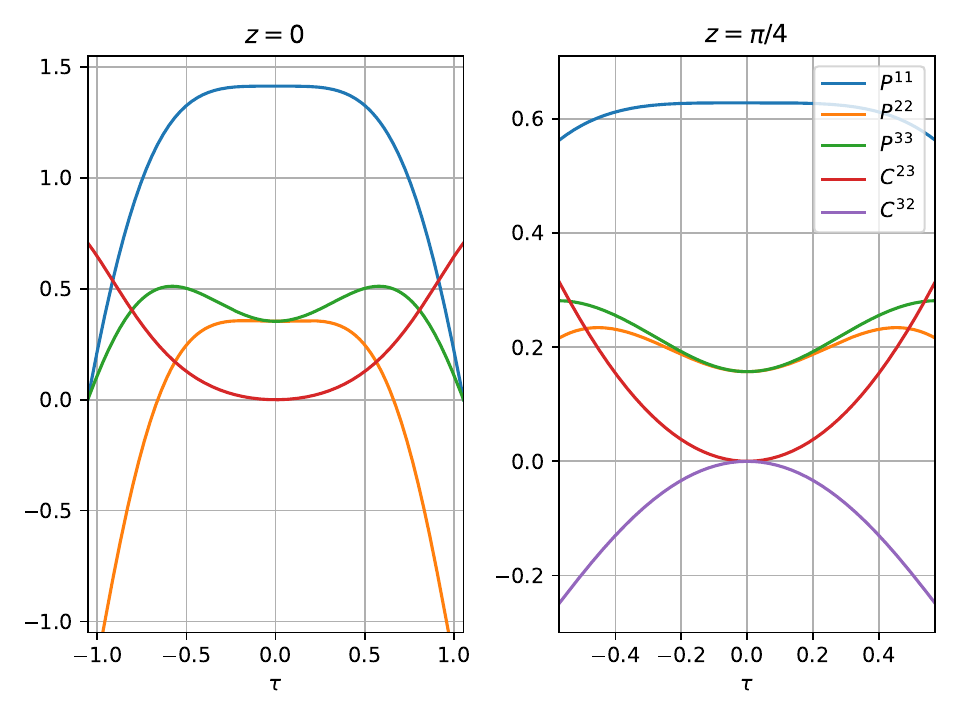}
    \caption{Non-vanishing AHS variables for the gravitational-electromagnetic wave collision at two different points of space, $z=0$ (left) and $z=\pi/4$ (right). Note that $C^{23}$ and $C^{32}$ coincide at $z=0$. It can be seen that all of the variables are regular at the singularity.}
    \label{fig: gw-em ahs behav}
\end{figure}

The spatial metric on a surface of constant $\tau$ is
\begin{equation}
q =e^{-M}\d z^{2}+e^{-U}\left(e^{V}\d x^{2}+e^{-V}\d y^{2}\right).
\end{equation}
We work with the triad
\begin{subequations}
\begin{align}
E_{1}^a & =e^{-U}\delta_{z}^a,\\
E_{2}^a & =e^{-M/2}e^{-U/2}e^{-V/2}\delta_{x}^a,\\
E_{3}^a & =e^{-M/2}e^{-U/2}e^{V/2}\delta_{y}^a.
\end{align}
\end{subequations}
With this choice, the non--vanishing AHS variables are
\begin{subequations}
\begin{align}
P^{11} &= -\frac{e^{-U}}{2\tau} U_\tau, \\
P^{22} &= -\frac{e^{-U}}{4\tau} \left(M_\tau + U_\tau + V_\tau \right), \\
P^{33} &= -\frac{e^{-U}}{4\tau} \left(M_\tau + U_\tau - V_\tau \right), \\
C^{23} &= \frac{1}{2}e^{-U} \left(U_z - V_z \right), \\
C^{32} &= -\frac{1}{2}e^{-U} \left(U_z + V_z \right).
\end{align}
\end{subequations}
Inserting our known solution Eq. (\ref{eq: metric component functions em-gw})
and inspecting term by term, we see that
\begin{subequations}
\begin{align}
U_\tau & = -\sqrt{2}\tau e^U \left(\sin 2u + \sin 2v \right),\\
V_\tau & =2\sqrt{2}\tau\frac{\cos u}{1+\sin u},\\
M_\tau & =-\sqrt{2}\tau\left(\tan u+\tan v\right) + V_\tau - \frac{1}{2} U_\tau,\\
U_z & = \frac{e^U}{\sqrt{2}} \left(\sin 2u - \sin 2v \right),\\
V_z & = -\sqrt{2}\frac{\cos u}{1+\sin u},
\end{align}
\end{subequations}
where $u$ and $v$ are understood to be functions of $\tau$ and $z$.
We therefore observe that in the interior where $0<u,v<\pi/2$,
all of the variables $P^{ij}$ and $C^{ij}$ remain finite and differentiable, including the singularity $u+v=\pi/2$. Fig.~\ref{fig: gw-em ahs behav} shows the AHS variables for the gravitational-electromagnetic wave collision as functions of $\tau$ at two sample points.

As a final remark, we note that as one approaches the singularity, the $C^{ij}$ and the triad $E^a_i$ vanish in this case as well, just as in the Kasner case.

\section{Discussion}
\label{sec: discussion}

The well-behavedness of the AHS variables at the singularities of the FLRW spacetimes, Bianchi I and IX models and the Schwarzschild metric had already been demonstrated. Here we have shown this to be the case for two specific colliding wave solutions. Although our results are for these two examples only, this further motivates the expectation that these variables are well behaved at generic spacelike singularities.

The first example we have looked at is the Khan-Penrose metric, which describes the collision of two impulsive gravitational waves. In order to calculate the AHS variables, we have chosen a foliation adapted to the singularity of the metric. Our results show explicitly that the AHS variables are finite at the singularity and can be continued through it without any problem.

The second example we have looked at is an exact solution which describes the collision of a gravitational wave with an electromagnetic wave. Here, too, are the AHS variables regular on the entire spacetime, including the singularity.

We should emphasize that since the singularities of general relativity occur at scales where quantum effects should be dominant and thus the classical theory is not necessarily a good approximation, investigating these singularities classically---as we have done---should be seen as a first step towards a fuller understanding of the problem. In fact, as we mentioned previously, there are a number of results indicating that quantum corrections become meaningful only when curvature or matter density are about a percent of the Planck scale. Therefore, it is possible that the classical results such as the BKL conjecture or the behaviour of the AHS variables at the singularity reliably describe what effectively happens near the singularity.

Of course, a full understanding of the problem requires a consistent theory of quantum gravity. However, semi-classical analyses such as looking at quantum fields on classical backgrounds \cite{Ashtekar2021} may also shed some light on the issue. This is what is referred to as ``level 2" in \textcite{ashtekar2022} and has been done, e.g., for cosmological models, and extending these semi-classical methods to colliding gravitational waves seems to be a reasonable way to gain further insight into their singularity structures.

\acknowledgements
K.\.I.\"U. is supported by T\"UB\.ITAK project no. 122F097. T.D. is partially supported by the Turkish Academy of Sciences (T\"UBA). We thank Fethi M. Ramazano\u{g}lu for many valuable suggestions.

\end{document}